\documentclass[aps,prb, twocolumn,amsmath,amssymb, superscriptaddress,citeautoscript,floatfix]{revtex4-2}
\usepackage{amsbsy}
\usepackage{wasysym}
\usepackage{bm}
\usepackage{color}
\usepackage{textcomp}
\usepackage{tikz-feynman,contour}
\tikzfeynmanset{compat=1.1.0}
\tikzfeynmanset{/tikzfeynman/momentum/arrow shorten = 0.3}
\tikzfeynmanset{/tikzfeynman/warn luatex = false}
\usepackage{gensymb}
\usepackage[breaklinks,colorlinks = true,linkcolor = red,urlcolor=cyan,citecolor=red]{hyperref}
\usepackage{array}
\usepackage{float}
\usepackage{graphicx}
\usepackage{subfigure}
\usepackage{float}
\usepackage{soul}

\usepackage{comment}
\usepackage{dcolumn,multirow}

\newcommand{\lvc}[1]{\unskip\ignorespaces}

\begin{document}

\title{
Topological  Impurity Bands
}

\author{Arnab Seth}
\affiliation{School of Physics, Georgia Institute of Technology, Atlanta, GA 30332, USA}
\author{Itamar Kimchi}
\affiliation{School of Physics, Georgia Institute of Technology, Atlanta, GA 30332, USA}

\date{September 8, 2026}

\begin{abstract}
The effects of disorder on topological phases of matter are typically either preservation of topology or its destruction. For example, the quantum Hall effect's quantization and chiral edge mode remain robust until a strong-disorder percolation transition exits the topological phase. Disorder can also enlarge a preexisting topological phase, yielding a topological Anderson insulator (TAI). However, the reliance of TAI and other theoretical treatments on perturbatively averaging out long-wavelength randomness suggests that additional effects might appear when the disorder involves localized defects. Here we show that a finite density of randomly distributed defects can generate various new topological phase transitions. This route to disorder-induced topology can be viewed as the formation of topological impurity bands (TIB) and is applicable even if the initial clean model is a trivial band insulator. The theory relies on an orbital-dependent vorticity of impurity bound states, which we derive analytically near a generic Dirac cone phase transition as well as using the Chern insulator lattice models of Qi-Wu-Zhang and Haldane. Their numerically computed phase diagrams at 3\% defect density show robust topological transitions in distinct TIB and TAI regimes, suggesting a local route to topological phases.
\end{abstract}

\maketitle

\clearpage

\begin{figure*}[t]
    \centering
    \includegraphics[width=1\linewidth]{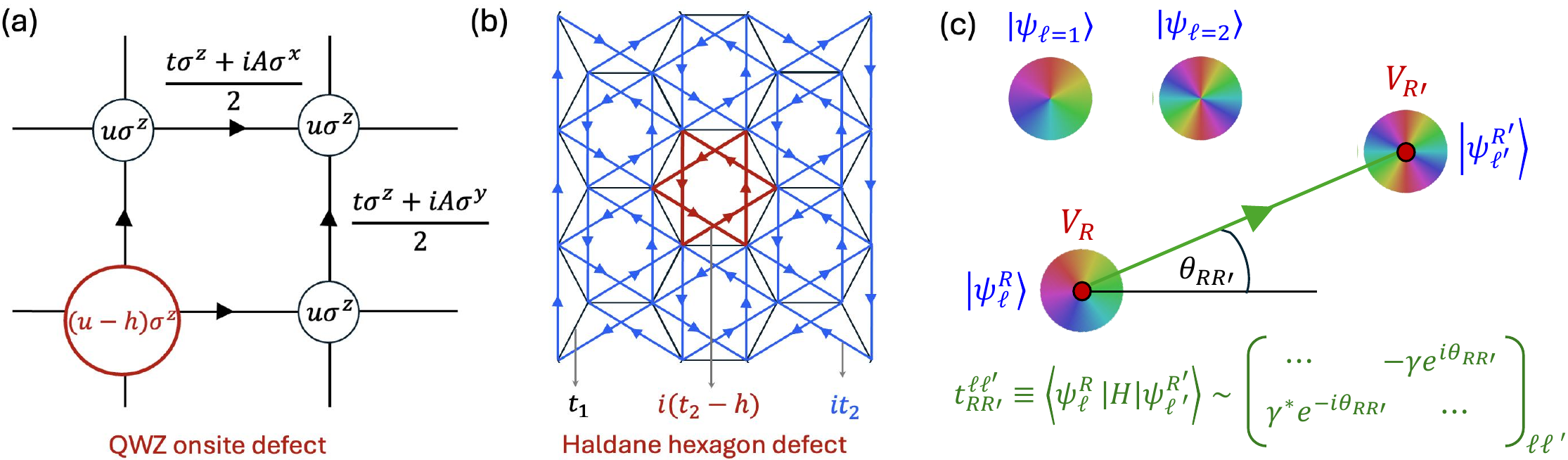}
    \caption{{\bf Impurities in  Chern insulators or trivial insulators and formation of topological impurity bands (TIBs). }
    (a)   Qi-Wu-Zhang (QWZ) model with onsite defect, Eq.~\ref{eq_defect_qwz}.
    (b) Haldane model with hexagon defect, Eq.~\ref{eq_defect_haldane}, capturing  spin-orbit coupling to a magnetic moment.  
    (c)  
    Each $V_R$ defect produces a pair of in-gap impurity bound states  $|\psi^R_\ell\rangle$  with differing vorticity $\ell=1,2$.
    When the states hybridize at finite impurity density they generate an effective low energy theory $t^{\ell \ell'}_{R R'}$  with QWZ-like  direction-dependent off-diagonal complex hopping $e^{\pm i \theta}$, producing TIB topological  phase transitions.
    }
    \label{fig_1}
\end{figure*}

The interplay of disorder and topology has long been a central problem in condensed matter physics. 
Topology is generally quite robust to disorder, as can be seen  in the paradigmatic case of the quantum Hall effect \cite{aoki_effect_1981, halperin_quantized_1982, chalker_percolation_1988}.
Even when local energy randomness closes the global spectral gap, and even when 2D disorder would lead to Anderson localization, the topological phase is not destroyed. Its quantized Hall conductivity and chiral edge modes arise from extended states that are topologically protected;
only when strong disorder eliminates these extended states, such as with a bulk percolation transition, does the system exit the topological phase \cite{aoki_effect_1981, halperin_quantized_1982, chalker_percolation_1988}. 

Disorder can also expand the phase boundaries of a preexisting topological phase, with the resulting disorder-induced phase transition known as the topological Anderson Insulator (TAI) transition~\cite{li_topological_2009,groth_theory_2009,prodan_threedimensional_2011}.
The TAI arises when the disorder-averaged self-energy $\langle \Sigma\rangle$ renormalizes the clean Hamiltonian $H_0$ such that the effective on-average Hamiltonian $H_0+\langle\Sigma\rangle$ lies within the topological phase.
Analogous TAI transitions occur in a variety of crystalline and amorphous systems 
\cite{guo_topological_2010, agarwala_topological_2017, meier_observation_2018, stutzer_photonic_2018, chen_realization_2024} through related mechanisms. 
The theories typically treat the effects of disorder perturbatively and on-average, with effects arising through longer wavelength inhomogeneities.
This suggests that a different theoretical approach may be necessary to fully characterize the effects of disorder when it involves local defects or impurities, as in the technologically relevant case of impurity doping.

An impurity-based theoretical approach has been productively used for the special case of   Yu-Shiba-Rusinov impurity states in superconductors \cite{balatsky_impurityinduced_2006}. In these superconducting systems magnetic impurities can transform a conventional $s$-wave superconductor into a topological superconductor. Realizations rely on helical impurity magnetism or on ferromagnetic impurities and strong spin-orbit coupling in the superconductor
\cite{pientka_topological_2013,weststrom_topological_2015,
brydon_topological_2015,nadj-perge_observation_2014}.
When the impurities are perfectly periodically arranged, producing a 2D Shiba lattice, the resulting superconductor topology is encoded as a Chern number of the impurity miniband 
\cite{rontynen_topological_2015,li_twodimensional_2016,soldini_twodimensional_2023}.
Randomly distributed impurities are less studied but show related effects \cite{poyhonen_amorphous_2018}.

Whether analogous physics from local impurities can arise in  generic band insulators remains largely unexplored and is a central motivation of our work. 
Impurities in gapped systems can produce localized in-gap bound states which gain additional protection if the system is topological \cite{gonzalez_impurity_2012,sau_bound_2013,slager_impuritybound_2015,kimme_existence_2016,diop_impurity_2020,michel_bound_2024, queiroz_ring_2024}. At sufficient impurity density or in alloys these bound states hybridize to form impurity bands. 
However, the ingredients that enable such impurity bands to acquire nontrivial topology in generic insulating systems have remained unknown. 

Here we identify a generic route for generating topological phase transitions by adding defects or impurities to a band insulator. The theory we derive becomes controlled in the limit of isolated impurities, where impurity bound states are precisely defined and the resulting topological phases can be viewed as arising from topological impurity bands (TIBs). 
The essential ingredient underlying TIB formation is a pair of impurity bound states that become degenerate and show nontrivial vorticity (phase winding) around the impurity, with differing vorticities for the two states. 
When hybridized these states yield a multi-orbital effective hopping model  with inter-orbital direction-dependent complex-amplitude hopping terms
that produce
topological phase transitions
and associated quantum anomalous Hall effects.

We establish this using a generic gapped Dirac cone field theory describing either the topological or trivial phase near a clean $H_0$ phase transition, as well as in two standard 2D lattice model examples: the Qi-Wu-Zhang (QWZ) \cite{qi_topological_2006}  and the Haldane honeycomb \cite{haldane_model_1988}  models (Fig.~\ref{fig_1}), again in both the topological and trivial phases of the $H_0$ model's phase diagram. The two models showcase two distinct microscopic mechanisms for generating vorticity.
Numerically studying the models (Fig.~\ref{fig:bott_ipr}) we find that even a small 3\% defect density is sufficient for generating various topological phase transitions  visible in the Bott index (which generalizes  Chern number to disordered systems \cite{loring_disordered_2011}). 
The transitions do not rely on the superlattice defect arrays 
considered previously \cite{kimme_existence_2016} and remain visible upon disorder averaging, or upon  changing the occupancies of states 
(as with nonzero temperature)
at energy scales
of order the $H_0$ band gap.
The numerically computed phase diagrams show TAI as well as TIB phase transitions, which match the corresponding analytical predictions and are also sharply distinguished by the inverse participation ratio (IPR) of low-energy states. This distinction suggests that TIBs arising from a finite density of local Hamiltonian modifications are a generic route to  controlling topological behavior through local effects.

\subsection{In-gap bound  states of a single impurity}

We begin by describing the 
 T-matrix approach to the in-gap bound  states of an isolated impurity. 
Starting with a clean band insulator  $H_0$ 
we introduce  a single impurity at location $R$ by the term  $V_R$, yielding   $ H=H_0+V_R$. 
The modification of the $H_0$ spectrum by the impurity can be computed in Brillouin-Wigner perturbation theory  around any particular $H_0$ frequency $\omega$.  We choose $\omega=0$ in order to identify in-gap states near zero energy.
The resummed perturbation theory, equivalent to including all scattering events from the single impurity, produces the scattering T-matrix whose poles give the exact renormalized spectrum:
\begin{align}
    T(\omega=0)=V_R\left(1-G_0V_R\right)^{-1},
\end{align}
where $G_0=-H_0^{-1}$ denotes the clean $H_0$ Green's function evaluated at zero energy. 

We will be interested in the wavefunctions of the associated bound states. 
Using $|\psi^{R}\rangle$ to denote the zero energy bound state which produces the T-matrix pole, we find it useful to decompose it into the local part which overlaps the $V_R$ impurity and the extended part away from $R$:
\begin{align}
    |\psi^R\rangle=|\psi^{V_R}\rangle+|\psi^{\perp_R }\rangle.
    \label{eq_splitting}
\end{align}
The impurity-projected state  $|\psi^{V_R}\rangle$  is defined by the local projector $P$  formed by the eigenvectors of $V_R$ with nonzero eigenvalues. Noting that $P V_R = V_R = V_R P$ and $P^2=P$, we find this term satisfies the implicit equation
\begin{align}
    |\psi^{V_R}\rangle \equiv P|\psi^R\rangle = (PG_0P) V_R|\psi^{V_R}\rangle,
    \label{eq_state_D}
\end{align}
which
also yields the critical impurity strength required for the zero-energy crossing of the impurity bound state. %
Once the impurity-projected part $|\psi^{V_R}\rangle$ is known, the remainder $|\psi^{\perp_R}\rangle$  is easily obtained by applying $G_0$ 
(which is still exponentially localized due to the $H_0$ gap)
\begin{align}
    |\psi^{\perp_R }\rangle\equiv (1-P)|\psi^R\rangle&=(1-P)G_0V_R|\psi^{V_R}\rangle
    \label{eq_state_Dabr}.
\end{align}

\subsection{Gapped Dirac cone with local mass defects} %

Surprisingly, under quite general conditions the zero energy bound state $|\psi^{R}\rangle$ can gain nontrivial vorticity. The vorticity can arise through at least two distinct mechanisms. As we now show, the first mechanism can be understood at the field theory level by considering $H_0$ to be a 2D Dirac cone with a (uniform) mass term.  
Since a single Dirac cone describes a topological transition that changes Chern number by 1, the gapped Dirac cone theory $H_0$ can capture either a trivial insulator or a topological (Chern) insulator phase. 
To see the impurity induced topology we will add local mass terms which locally modify the Dirac cone uniform mass, and show how such a defect produces a pair of in-gap impurity bound states with different nonvanishing vorticities.

The uniform gapped Dirac cone with velocity $v$ is
\begin{align}
    H_0=v\left(\sigma^x q_x+\sigma^yq_y\right)+\frac{\Delta}{2}\sigma^z.
    \label{eq_dirac}
\end{align}
The mass term $\Delta/2$ is odd under time reversal (TR) (with $\hat{\Theta}^2=-1$) 
but preserves the model's particle-hole symmetry  $\hat{\mathcal{P}}^2=1$.
Note that though this Dirac model is in symmetry class D it captures many low energy transitions since
every symmetry class of topological insulators
has a massive Dirac Hamiltonian representative in the same topological class \cite{ryu_topological_2010}. 
Here we have
$\hat{\mathcal{P}}=\sigma^x\mathcal{K}$ and $\hat{\Theta}=i\sigma^y\mathcal{K}$ with $\mathcal{K}$ complex conjugation.

Now consider a generic local impurity $V_R$  which, just like $\Delta$ above, breaks $\hat{\Theta}$ but respects  $\hat{\mathcal{P}}$. %
These symmetry considerations restrict the impurity to a $\sigma^z$ mass term. For concreteness we take %
\begin{align} 
    V_R= -h \Theta(r_0-r)\sigma^z
\end{align}
where the Heaviside step function $\Theta$ controls the impurity radius $r_0$.  %
Since $\hat{\mathcal{P}}^\dagger V_R\hat{\mathcal{P}}=-V_R$  the impurity spectra is always symmetric about zero, and without loss of generality  it has  two low energy eigenstates with energies $\pm E$, related by $\hat{\mathcal{P}}$. 
Since $V_R$ is odd under TR the two states also form a Kramers doublet.

To study the in-gap zero-energy bound states of a single impurity we simply use the impurity projected T-matrix; see the Methods section for details. The impurity projected Green's function is 
\begin{align}
    PG_0(r)P\propto \sigma^z    
    \label{eq_green_dirac_projected}
\end{align}
which  multiplies with $V_R\propto \sigma^z$ to give the identity, $PG_0(r)PV_R\propto \left(\sigma^z\right)^2=\mathcal{I}$. Hence solving Eq.~\ref{eq_state_D} enables a degenerate pair of impurity bound states, $\psi^{R}_{\ell}$ with $\ell=1,2$. This emergent-orbital index $\ell$ is associated with vorticity, as seen from the Eq.~\ref{eq_state_Dabr} extended wavefunctions
\begin{align}
    \psi^{\perp_R}_{1} \sim \left(1, -i e^{i\theta}\right)^T , \ \ 
    \psi^{\perp_R}_{2}
    \sim \left(i e^{-i\theta},1\right)^T 
    \label{eq_dirac_extended}
\end{align}
with $\theta$ the polar coordinate far from the impurity.

This pair of bound states associated with each impurity provides a low-energy manifold. 
The states cross each other (become degenerate) at a critical value of the impurity strength $h_c$ which depends on the lattice regularization, as we will see explicitly in the next two sections by considering two lattice model realizations. 
After analyzing the two lattice models we will turn to the question of multiple impurities, where these two (or more) low-energy states will serve a key role. 
When impurities occur at finite density such that the hybridization between  states of neighboring impurities is substantial, they yield an effective low-energy theory (Fig.~\ref{fig_1}(c) and Eq.~\ref{eq_heff}) with a generalized amorphous QWZ structure that can produce topological phase transitions and thereby form a TIB.

\subsection{QWZ model with onsite defects}

As an explicit lattice realization of the gapped Dirac Hamiltonian   we consider  the QWZ model \cite{qi_topological_2006}, which  realizes $C=0,\pm1$ trivial and Chern insulators  on the square lattice with two onsite orbitals $l$. 
The clean QWZ model $H_0$ is
\begin{align}
    H_0&=u\sum_i  \sigma^z_{ll}\hat{n}_{i,l}+\frac{1}{2}\sum_{\langle ij\rangle}\left(c^\dagger_{il}T_{ij}^{ll'}c_{jl'}+{\rm H.c.}\right)\\
    T^{ll'}_{ij}&=\left(\begin{array}{cc}
       t &  iAe^{-i\theta_{ij}}\\
       iAe^{i\theta_{ij}}  & -t
    \end{array}\right)_{l l'}
    \label{eq_qwz}    
\end{align}
with the number operator $\hat{n}=c^\dagger c$ and a sum over repeating orbital indices $l$ implied. We then add onsite TR-breaking impurities $V_R$  at sites $R$  that act on the two orbitals by
 \begin{align}
    V_R=-h ~
    \sigma^z_{ll}\hat{n}_{R,l}.
    \label{eq_defect_qwz}
\end{align}
Positive $h$ reduces   $u$ locally; see Fig.~\ref{fig_1}. Here $\theta_{ij}$ is the polar angle of  $\vec{r}_j-\vec{r}_i$ measured from the $x-$axis, consequently $\theta_{ji}=\theta_{ij}+\pi$. 
With $t>0$, this QWZ model shows trivial $C=0$ phases for $|u|>2t$, and topological phases $C=-1$ for $-2t<u<0$ and $C=1$ for $0<u<2t$. 
Each $|u|=2t$ phase transition is governed by the  single Dirac cone low energy theory, with velocity $v=A$ and mass term  $\Delta/2={\rm sgn}(u)(|u|-2t)$.  The Dirac cone is at the $\Gamma$  point near $u=-2t$ and $M$ point near $u=2t$, which produces  additional unimportant sign oscillations. %

We can now compute the bound states and the critical impurity strength required for their zero energy crossing. Here  $P=\sum_\mu|{ R}\mu\rangle \langle { R}\mu|$ is an onsite projector, giving
\begin{align}
    PV_RP=-h\sigma^z~~ ,  ~~PG_{0}(r)P=g_{\rm onsite}\sigma^z    
\end{align}
with $g_{\rm onsite}=-(2\pi)^{-2}\int d^2k \left(u+t\left(\cos k_x+\cos k_y\right)\right)/\varepsilon_{k}^2$ being the lattice regularized %
onsite Green's function, and  $\varepsilon_{k}$  the dispersion  of $H_0$; see Methods. 
As in the Dirac  case these two  $\sigma^z$ matrices  multiply to the identity giving two degenerate orbitals $|\psi^{R}_{\ell}\rangle$ with differing vorticity $\ell$.
The critical impurity strength where the in-gap states cross zero energy is  obtained by setting the eigenvalues of 
$(-h_c\sigma^z)(g_{\rm onsite}\sigma^z)=-h_c g_{\rm onsite}\mathcal{I}$ to 1, which 
 gives 
\begin{align}
    h_c=-g_{\rm onsite}^{-1}.
    \label{eq_hc_qwz}
\end{align}

The extended part of the wavefunction can be obtained as before by applying the QWZ Green's function on $|\psi^{V_R}_{\ell}\rangle$. 
Since the long range part of the QWZ Green's function is well approximated by the low energy Dirac theory we obtain the same vorticity structure as in Eq.~\ref{eq_dirac_extended}.

\subsection{Haldane model with hexagon defects}

The vorticities of the pair of zero-energy in-gap bound states can also arise via a  second, distinct, mechanism, independent of $G_0$. 
Consider an impurity $V_R$ extending beyond a single lattice site and preserving a rotation symmetry. If the impurity-free Hamiltonian $H_0$ also possesses the same rotation symmetry then the impurity bound states should carry the symmetry quantum numbers, namely angular momentum,  producing vorticity.

To show this second mechanism we consider the  Haldane honeycomb model \cite{haldane_model_1988}, which can realize  $C=0,\pm1$ trivial and Chern insulator phases on the honeycomb lattice with imaginary second neighbor hoppings. 
We then introduce a  TR-breaking defect $V_R$ that modifies these second neighbor hoppings on a particular hexagon $\hexagon'$. These could arise through the same mechanism that generates the uniform $t_2$, for example via Kane-Mele spin-orbit coupling \cite{kane_quantum_2005} written for each spin species or upon integrating out a magnetic moment at the hexagon center.  So in this case we have the clean  Hamiltonian $H_0$  
\begin{align}
        H_0&=-t\sum_{\langle ij\rangle}c_i^\dagger c_j+i t_2\sum_{\hexagon}\sum_{\langle\langle ij\rangle\rangle\in \hexagon}c_i^\dagger c_j\nonumber\\
        &\hspace{1.5cm}+\frac{m}{2}\sum_{i}\left(c^\dagger_{iA}c_{iA}-c^\dagger_{iB}c_{iB}\right) +\text{H.c.}, 
\end{align}
with defects added at hexagon positions $R$ via
\begin{align}
       V_R&=-i h\sum_{\langle\langle ij\rangle\rangle\in \hexagon'}c_i^\dagger c_j+\text{H.c.}.
       \label{eq_defect_haldane}
\end{align}
Positive $h$ reduces   $t_2$ locally. The second neighbor hoppings $ \langle\langle ij\rangle\rangle\in \hexagon $ are always traversed counterclockwise around each hexagon; see Fig. \ref{fig_1}.

Note that although the low energy theory of $H_0$ near its transitions ($m=3\sqrt{3}t_2$) can again involve gapped Dirac cones, 
there is an important distinction. In this case   $H_0$ has a threefold rotation $C_3$ symmetry around every site and hexagon center,
which prevents the Green's function from gaining a phase winding, in contrast to QWZ. 
However, this same $C_3$ symmetry of $H_0+V_R$ around the $\hexagon'$ center $R$ (which is further enhanced to a $C_6$ symmetry in the sublattice symmetric limit of $m=0$) enables the impurity bound states to exhibit nontrivial angular momentum and associated vorticity.

This occurs as follows; see Methods for derivation details.
Following Eq.~\ref{eq_state_D}, the critical impurity strength $h_{c}$ required for the zero energy crossing is obtained by   finding $(PG_0P)V_R$ and setting its eigenvalues  to  1. This gives
\begin{align}
h_{c,\beta}^{-1}=-3t_2\tilde{g}_{2ai}+\beta \sqrt{3m^2\left(\tilde{g}_{0m}{-}\tilde{g}_{2mr}\right)^2+3g_{31}^2}
    \label{eq_hc_haldane}
\end{align}
with $\tilde{g}_{2ai}$, $\tilde{g}_{0m}$, $\tilde{g}_{2mr}$, and $g_{31}$ given in Methods.
Zero crossings occur at two critical points, denoted by $\beta=\pm$.
At each critical point there are two degenerate zero energy bound states, computed as the eigenvectors of $(PG_0P)V_R$.
Parameterizing the two states by an additional index $\alpha=\pm$, or equivalently by $\ell=(3-\alpha\beta)/2$ with $\ell=1,2$, 
these bound states are given by
\begin{align}
    \psi^{\perp_R}_{\ell}({\vec r}, S)=%
    g^S_{\beta\alpha}(\rho)e^{i\beta \ell \phi}.
    \label{eq_dbar_mzero}
\end{align}
Here $S$ denotes sublattices $A,B$ and $\vec{\rho}=({\vec r}-{\vec R})/\xi$ with $\phi$ its polar angle and $\rho = |\vec{\rho}|$. The radial function $g^S_{\beta\alpha}(\rho)$ is a real 
function whose sign is independent of the sublattice $S$ and magnitude decays exponentially  away from the impurity. 
In the particle-hole symmetric limit of $m=0$ we further get $g^A_{\beta\alpha}(\rho)=g^B_{\beta\alpha}(\rho)$ implementing the enhanced $C_6$ symmetry.
The phase windings $e^{i\beta \ell \phi}$ arise already in  $\psi^{V_R}_{\ell}$ and are unchanged in $\psi^{\perp_R}_{\ell}$.

These $\ell=1,2$ single and double vorticities, with either $\beta$ sign, give the four distinct TR-breaking angular momenta of $C_6$ rotations. Importantly, they arise here even when $C_6$ symmetry is broken down to $C_3$ by a sublattice imbalance mass term $m$; see Methods. %

\begin{figure*}
    \includegraphics[width=1\linewidth]{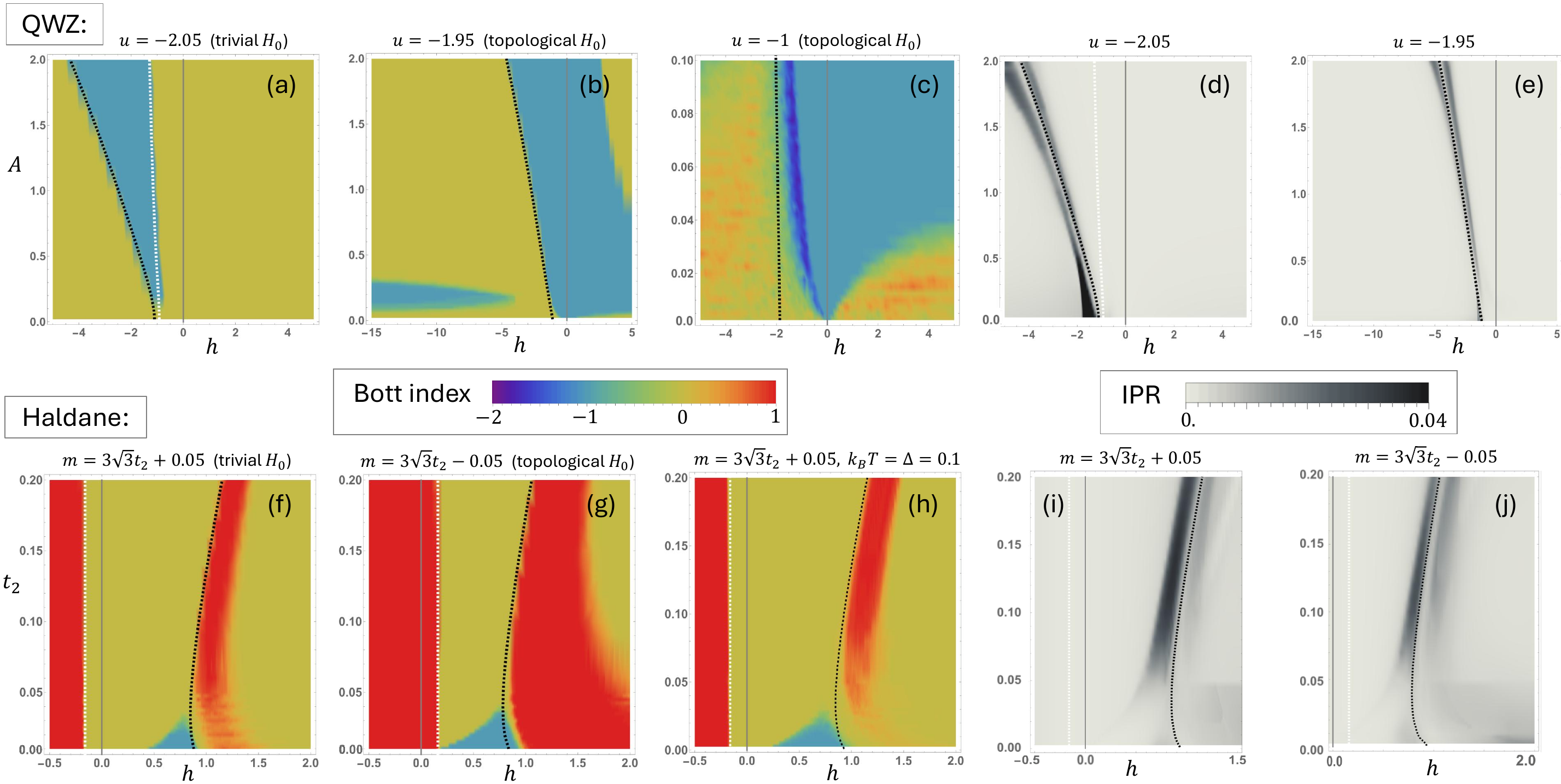}
    \caption{{\bf 
    Numerical evidence for TIB and TAI 
    phase transitions for various $H_0$ Hamiltonians. 
    }
    3\% impurities per site were randomly placed in a %
    1024-site system in 25 disorder configurations,
    with the impurity-free model $H_0$ being QWZ (top row) or Haldane (bottom row).
    In each panel horizontal axis is defect strength $h$ (where $h=0$ is the clean model $H_0$) and vertical axis is an $H_0$ parameter ($A$ for QWZ, $t_2$ for Haldane). 
    The disorder averaged Bott index (color scale) shows clear topological phase transitions %
    in both the trivial (a,f) and topological (b,g) phases of the clean model $H_0$, with $H_0$ gap  $\Delta=0.1$ near its  Dirac cone phase transition ($u=-2$ or $m=3\sqrt{3}t_2$), as well as deep in the QWZ topological phase (c) with  $\Delta=2A$.
    (The Haldane model $m$ was varied with $t_2$ to preserve $\Delta=0.1$.)
    The analytically computed T-matrix  bound-state zero-crossing $h_c$ results (black dotted lines), which are controlled at small defect density, agree well with the corresponding TIB transitions. 
    The analytical estimates for disorder-averaged TAI transitions (white dotted lines), which as expected are nearly $h$-independent (vertical), also agree with the numerics. (No TAI solutions occur in b,c.)
    The distinction between TAI and TIB transitions is also observed in 
    the disorder-averaged inverse participation ratio (IPR, shown in greyscale), which peaks near $h_c$ transition lines as expected from impurity bound states, while vanishing at TAI transitions (panels d,e,i,j corresponding to a,b,f,g respectively).
    Finally, note that for many parameters Bott indices are unchanged when state occupancies are modified by a Fermi-Dirac distribution at temperature $k_B T=\Delta=0.1$  (panel h, with $H_0$ as in f).
    }
    \label{fig:bott_ipr}
\end{figure*}

\subsection{Amorphous QWZ-type effective low-energy model}

The effects  of the in-gap states' vorticities become important when impurities occur at finite density and the bound states from different defects hybridize. As we now show, thanks to the phase winding terms, the 
effective Hamiltonian generated for the low energy (in-gap) subspace gains QWZ-type vorticity terms that enable it to produce Chern-number-changing topological phase transitions. 
\label{sec:lowEmodel}

We analytically compute the hybridization of the impurity bound states
near their zero-energy crossing i.e.\ for $h$ near $h_c$, which allows the bulk states to be integrated out.
The hybridization leads to an effective low energy hopping model on the defect sites $R$. This is the effective low energy Hamiltonian which can produce topological phase transitions through the impurity states, as we now derive.
Defining the fermionic creation ($a^{\dagger}_{R,\ell}|0\rangle \equiv|\psi^R_{\ell}\rangle$) and annihilation %
operators for the zero-energy impurity bound states in orbital $\ell$, we can write the effective hopping model as
\begin{align}
    \mathcal{H}^{\rm eff}_{\rm imp}&=(h-h_c)\sum_{ R}\sigma^z_{\ell\ell}\hat{n}_{R,\ell}+ \sum_{R, R'}t^{\ell\ell'}_{RR'}a^{\dagger}_{R,\ell} a_{R',\ell'},
    \label{eq_heff}
\end{align}
where $\hat{n}=a^\dagger a$ is the number operator, $R,R'$ are the positions of the defects, and sums over $\ell$ and $\ell'$ indices are implied.  

The first term is an onsite orbital splitting term (or chemical potential imbalance) for the zero energy states which  arises as we deviate from $h_c$, since the two states no longer remain at zero energy. 

In the second term $t^{\ell\ell'}_{RR'}$ denotes the overlap integral of the two zero energy bound state wavefunctions at  $R$ and $R'$ due to the  two impurity Hamiltonian $H=H_0+V_R+V_{R'}$ at those sites. This is given by
\begin{align}
    t^{\ell\ell'}_{RR'}&=\langle  \psi^R_{\ell}| H|\psi^{R'}_{\ell'}\rangle    =\langle\psi^{R}_{\ell}|V_{R'}|\psi^{R'}_{\ell'}\rangle
    \\
    &=\langle \psi^{V_R}_{\ell}|V_RG_0(R-R')V_{R'}|\psi^{V_{R'}}_{\ell'}\rangle.
    \label{eq_overlap_with H}
\end{align}
The expressions are obtained using %
$(H_0+V_R)|\psi^R_{\ell}\rangle=0$ (similarly for $R'$) %
followed by
the extended $\psi^{\perp_R}_{\ell}$ formula (Eq.~\ref{eq_state_Dabr}), see Methods. The associated onsite term decays much faster and is approximated to $t^{\ell\ell'}_{RR}\approx0$.

Computing Eq.~\ref{eq_overlap_with H}  for the two models we obtain
\begin{align}
    t^{\ell\ell'}_{RR'}=h^2\left(
    \begin{array}{cc}
       s^{11}_{RR'}  & -\gamma s^{12}_{RR'}e^{i\beta\theta_{RR'}} \\
       \gamma^*s_{RR'}^{21}e^{-i\beta\theta_{RR'}}  & -s^{22}_{RR'}
    \end{array}
    \right)_{\ell\ell'}.
    \label{eq_eff_hopping}
\end{align}
For QWZ, $\gamma=-i$ and $\beta=1$; 
for Haldane, $\gamma=1$, and $\beta=\pm1$ corresponding to each $h_{c,\beta}$ critical point. 
(Note that $\gamma=-i$ and $\gamma=1$ are not gauge equivalent in a generic non-bipartite defect graph.)
The functions $s_{RR'}^{\ell\ell'}$ involve $G_0$, and for the Haldane case, a sum over defect sites computable via multipole expansion in defect size over defect separation; see Methods.
In both cases the functions contain no phase winding. They decay exponentially with the model correlation length; to observe $t^{\ell\ell'}_{RR'}$ physics the impurity states should be sufficiently extended (shallow) relative to their density such that the magnitude of $t$ between neighboring impurities is not exponentially small 
relative to other scales in the problem. %

The key features of $t^{\ell\ell'}_{RR'}$ are its inter-orbital direction-dependent complex hopping terms $e^{\pm i\beta\theta_{RR'}}$. Thanks to these terms the effective low energy model %
$\mathcal{H}^{\rm eff}_{\rm imp}$
can be viewed as an amorphous generalization of a QWZ type model. Such models are known to produce topological phase transitions just like their crystalline counterparts \cite{agarwala_topological_2017,dornellas_quantized_2022,roy_phase_2026}. 
The capacity of the vorticity-based inter-orbital term to generate topology has also been analyzed in terms of band inversions with relative angular momentum ~\cite{venderbos_higher_2018} which generate Chern numbers.

\subsection{\label{sec_numerics}Robustness of impurity induced topological transitions and TAI-TIB distinctions} %

To test these theoretical predictions we numerically computed the phase diagrams of Haldane and QWZ models with 
defects at density $n_d=3\%$  impurities per site.
Without translation symmetry  we rely on the disorder averaged Bott index~\cite{loring_disordered_2011} to characterize  topology. 
Particle occupancy is fixed to half filling.
Energy units are set by $t=1$. 
The results are shown in  Fig.~\ref{fig:bott_ipr}.

In panels a,b,f,g we focus on parameter regions  on either side of an $H_0$ Dirac cone phase transition,  with $H_0$ Dirac cone gap $\Delta=0.1$. This enables impurity states to hybridize strongly even at 3\% defect density.
Across most of these parameters the  disorder-averaged Bott index is well quantized.
The main finding is clear: adding disorder by tuning the impurity strength $h$ away from $h=0$ produces a variety of topological phase transitions, regardless of whether the $h=0$ clean model $H_0$ is in a trivial or topological phase.

Conventional on-average disorder effects lead to TAI transitions when the disorder-averaged $H_0+\langle\Sigma\rangle$ parameters produce a band gap closing. We compute this disorder correction analytically to order $h^2$ and find that the QWZ $u$ and Haldane $t_2$ parameters renormalize by
\begin{align}
    u\rightarrow u^{\rm eff}&=u-{n_d} h+n_d(1-n_d)g_{\rm onsite}h^2,
    \label{eq_tai_qwz}\\
    t_2\rightarrow t_2^{\rm eff}&=t_2-2n_d h.
    \label{eq_tai_haldane}
\end{align}
The TAI transitions then occur when $u^{\rm eff}=-2$ or $t_2^{\rm eff}=m/3\sqrt{3}$.
For the Haldane model, $t_2$ gains no $h^2$ corrections; though  other terms including $m$ gain $n_d h^2$ corrections, these can be neglected since the $t_2$ linear term already sets $|h_c|  \approx 0.16$.
The resulting TAI transitions are plotted as white dotted lines, which by necessity are nearly vertical: the on-average disorder effects that lead to TAI transitions are only weakly dependent on varying other $H_0$ parameters.

In contrast, the predicted TIB transitions show a remarkable richness. The bound state level crossing $h_c$ (black dotted line), analytically computed from the T-matrix in  Eqs. \ref{eq_hc_qwz} and \ref{eq_hc_haldane} for an isolated impurity, 
successfully predicts nearby phase transitions. 
Additional topological transitions also occur, as expected from Eq.~\ref{eq_eff_hopping}
though away from the $h\approx h_c$ controlled limit of the theory.

We also present results for $H_0$ far from its Dirac cone phase transitions, using the QWZ model with small gap $\Delta=2A$ in panel c.
The TIB transitions near $h\approx h_c$ persist. Here they capture transitions out of the Chern insulator with gapless regions exhibiting non-quantized disorder-averaged Bott indices.

The energy scales that protect each Chern insulator phase are not vanishingly small, as can be seen by comparing panels f and h. In panel h the Bott index computation was modified by replacing the projector to occupied states by the Fermi-Dirac distribution at a finite temperature $k_B T = \Delta = 0.1$. %
Though the Bott index, like Chern number, is only defined and quantized at zero temperature, this comparison is instructive: 
many of the transition lines remain visible at this finite temperature.

To further distinguish the TAI and TIB transitions we also compute the inverse participation ratio (IPR) of the highest occupied state, shown in panels d,e,i,j corresponding to a,b,f,g. As expected, the TAI transitions always show vanishing IPR, while  other transitions can show nonzero IPR, which peaks near the T-matrix $h_c$ lines. This IPR is due to low energy states clustering near impurity sites, confirming that the associated $h_c$ transitions involve strongly localized impurity bound states, even while other transitions, which may be governed by shallow impurities, show extended states.

\subsection{Outlook}

The orbital-dependent vorticity of in-gap states and resulting topological impurity bands provide a general framework for understanding the formation of topological phase transitions in systems with defects, dopants, alloys, or other local modifications at finite density.
Experimental platforms for realizing TIBs are quite varied and increasingly investigated, 
in both fermionic and bosonic systems,
ranging from 
topological photonic alloys \cite{qu_topological_2024}; 
to amorphous mechanical metamaterials \cite{mitchell_amorphous_2018}; to 
 Moire superlattices 
\cite{chen_tunable_2020}; 
to magnetically doped semiconductors such as chromium-doped (Bi,Sb)$_2$Te$_3$  \cite{chang_experimental_2013};
and to non-magnetic quantum spin Hall systems such as amorphous Bi$_2$Se$_3$
\cite{corbae_observation_2023}, 
which can be explicitly described with a QWZ (or sometimes Haldane) model for each spin species \cite{bernevig_quantum_2006, kane_$z_2$_2005}.
In addition to the 
quantized anomalous Hall effect and associated signatures of Chern insulators,
the TIB theory implies local vorticity effects 
which could  be directly observed in local probes via circulating currents or orbital magnetization. %

Several theoretical connections appear intriguing. Though we have framed our discussion in terms of the 2D Chern number, the theory should extend to other dimensions and symmetry classes.
There may be a relation to the fluctuating moments of Kondo topological insulators \cite{dzero_topological_2010} and to the local topology of classical spins \cite{michel_bound_2024}. 
Within the framework of ring states for topological phases \cite{queiroz_ring_2024}, the ring structure is obviously amenable to carrying vorticity, but note that the vorticity we compute does not rely on $H_0$ being topological.
Our consideration of local mass terms was inspired by associated effects from Stone-Wales lattice defects \cite{seth_chiral_2026} and our use of the impurity projected T-matrix was inspired by the chirality reversal recently discovered in gapless Dirac cones  \cite{neehus_genuine_2025,xu_chirality_2026}, with possible implications for the understanding of renormalization group flows of disordered Dirac fermions.

\subsection{Acknowledgments}

We thank Zhu-Xi Luo, Xueda Wen and Yahui Zhang for helpful discussions.
This work was supported by the U.S. Department of Energy, Office of Science, Basic Energy Sciences, under Early Career Award Number DE-SC0025478.

\subsection{Methods}

\subsubsection{Dirac cone computation.} 
To study each of the two zero energy bound states $
    |\psi^R\rangle$ we begin by using Eq.~\ref{eq_state_D}  
for the  impurity projected part $|\psi^{V_R}\rangle$, which entails diagonalizing $(PG_0P)V_R$. 
The real space zero energy Green's function $G_0(r)= -(2\pi)^{-2}\int d^2k~ H_0^{-1}e^{ik\cdot r}$ of the gapped Dirac Hamiltonian  is

\begin{align}
    G_0(r)&= (2\pi^2\xi^2\Delta)^{-1}\left(  K_0 \left(  r/\xi  \right) \sigma^z  -i  K_1\left(r/\xi\right) \hat{r}\cdot \vec{\sigma}\right)
\end{align}
where $\xi=2v/\Delta$ is the correlation length, and $K_0(x)$ and $K_1(x)$ are Modified Bessel functions  of the second kind. These functions decay exponentially at large distances as $\sim \sqrt{\frac{\pi \xi}{2 r}}e^{-r/\xi}$.
Note that the  second term of the Green's function is TR symmetric, but has a non-trivial winding under rotation about the $z$-axis. When the  Green's function is projected onto the impurity subspace,  the second term vanishes due to $\hat{\mathcal{P}}$ symmetry leading to Eq. \ref{eq_green_dirac_projected} (see SI).
The proportionality function is determined by  the details of the impurity eigenstates and the  short distance behavior of the Green function, whose divergence must be regularized. 
Now since $V_R\propto \sigma^z$ its product with $P G_0P$ is the identity, $PG_0(r)PV_R\propto \left(\sigma^z\right)^2=\mathcal{I}$, hence the two eigenstates of the impurity give rise to two degenerate zero energy poles of the T-matrix.
We label this degeneracy by an index $\ell=1,2$, which as we now show is associated with vorticity.

To compute the extended part of the zero energy bound state $|\psi^{\perp_R}\rangle$,  following Eq.~\ref{eq_state_Dabr} we  apply $G_{0}(r)$ on $|\psi^{V_R}\rangle$. The second term of $G_0$ produces a non-trivial phase winding,
\begin{align}
    &\psi^{\perp_R}_{1}({\vec r})\sim %
    \left(K_0(\rho), -i K_1(\rho)e^{i\theta}\right)^T\\
    &\psi^{\perp_R}_{2}({\vec r}) \sim %
    \left(iK_1(\rho)e^{-i\theta},  K_0(\rho)\right)^T
\end{align}
where ${\vec \rho}=({\vec r}-{\vec R})/{\xi}$ is the position vector measured from the impurity position in the units of  correlation length $\xi$,  and $\theta$ is the polar angle of ${\vec \rho}$.
Far from the impurity $\rho\gg1$ the asymptotic limit of the Bessel functions $K_{0,1}(\rho)\approx\sqrt{\frac{\pi}{2\rho}}e^{-\rho}$ 
can be pulled out, highlighting the $e^{\pm i \theta}$ vorticities.

\subsubsection{QWZ model computation.}

The QWZ model in momentum space is $H_0 = \sum_k H_{0,k}$ with $H_{0,k}$ given by
\begin{align}
    \left(u+t\left(\cos k_x+\cos k_y\right)\right)\sigma^z+A\left(\sin k_x\sigma^x+\sin k_y\sigma^y\right).
\end{align}
The dispersion $\varepsilon_k$ is set by
\begin{align}
    \varepsilon_k^2=\left(u+t\left(\cos k_x+\cos k_y\right)\right)^2+A^2\left(\sin^2 k_x+\sin^2 k_y\right)
\end{align}
The zero energy Green's function $G_0(r)$ has a diagonal part $g_{\rm 11}\sigma^z$ and off diagonal terms $g_{\rm 12}=-g_{21}^*$, with
\begin{align}
    g_{11}(r)&=-\frac{1}{4\pi^2}\int d^2k ~ e^{ik\cdot r}~\frac{u+t(\cos k_x+\cos k_y)}{\varepsilon_k^2}  \\
    g_{12}(r)&=-\frac{1}{4\pi^2}\int d^2k ~e^{ik\cdot r}~\frac{A(\sin k_x-i \sin k_y)}{\varepsilon_k^2} 
\end{align}
Note that $g_{\rm onsite}=g_{\rm 11}(r=0)$.

\subsubsection{Haldane model computation.}

The Haldane model in momentum space is $H_0 = \sum_k H_{0,k}$ with $H_{0,k}$ given by
\begin{align}
         \left(c^\dagger_{k,A}~~c^\dagger_{k,B}\right)
        \left(
        \begin{array}{cc}
           m+t_2 d_k  & h_k \\
            h_k^* & -m-t_2 d_k 
        \end{array}\right)
        \left(\begin{array}{c}
             c_{k,A}  \\
             c_{k,B}
        \end{array}\right)
\end{align}
\begin{align}
        &d_k=-2\left(\sin k_1+\sin k_2-\sin(k_1+k_2)\right),\\
        &h_k= -t(1+e^{i k_1}+e^{-ik_2}).
\end{align}
The dispersion is 
\begin{align}
    \varepsilon_k=\pm \sqrt{(m+t_2d_k)^2+|h_k|^2}
\end{align}

Now consider an impurity. 
To compute the impurity projected Green's function $G_0$ and defect bound state, note that the support of $V_R$ is formed by its eigenvectors with non-zero eigenvalues, which are given by:
\begin{align}
    V_R|v_\pm,S\rangle = \pm \sqrt{3} h |v_\pm,S\rangle.
\end{align} 
Here $\langle {\vec r}|v_{\alpha},S\rangle=\frac{1}{\sqrt{3}}e^{ i\alpha\phi_d}$ is the defect wavefunction on the three $S$-sublattice sites of the defect hexagon $\hexagon'$ with $\phi_d$ denoting the angle of the site ${\vec r}$ (which can be 0, $2\pi/3$ or $4\pi/3$ up to a reference angle) from the $\hexagon'$ center. Note that these wavefunctions are independent of sublattice $S$.
The defect projector $P$ is then given by  $P=\sum_{S=A,B}\sum_{\alpha=\pm} |v_\alpha,S\rangle\langle v_\alpha,S|$. 
To obtain $|\psi^{V_R}\rangle$ we need to find $(PG_0P)V_R$. Denoting Pauli matrices  acting on the $\alpha$ and $S$ indices as $\mu$ and $\sigma$, respectively, we find  
\begin{align}
    \frac{1}{\sqrt{3}h\mathcal{N}}(PG_0P)V_R = \ \  a_0\sigma^0\mu^0+\cos\theta_0 \sigma^z\mu^z
    \nonumber \\ 
    -\sin\theta_0 \cos\phi_0\sigma^x\mu^z
    +\sin\theta_0 \sin\phi_0\sigma^y\mu^0
    \label{eq_tmatrix_haldane}
\end{align}    
with $\sigma^0$ and $\mu^0$ being the identity matrices. Note that this matrix is still diagonal in the $\alpha=\pm$ space since only $\mu^0$ and $\mu^z$ appear. 
Here $\phi_0=2\pi/3$, $\cos\theta_0={m(\tilde{g}_{0m}-\tilde{g}_{2mr})}/{\mathcal{N}}$, $\sin\theta_0=-g_{31}/\mathcal{N}$, $\mathcal{N}=\left(m^2(\tilde{g}_{0m}-\tilde{g}_{2mr})^2+g_{31}^2\right)^{1/2}$, and $a_0=-\sqrt{3}t_2\tilde{g}_{2ai}/\mathcal{N}$.
The terms 
$g_{31}$, $\tilde{g}_{0m}$, $\tilde{g}_{2mr}$ are various Green's function elements in real space,
\begin{align}
    & \tilde{g}_{0m}=\frac{1}{4\pi^2}\int \frac{\eta_e+2t_2^2d_k^2}{\eta_e^2-\eta_o^2}d^2k,
    \\
    & \tilde{g}_{2mr}=\frac{1}{4\pi^2}\int \frac{\eta_e+2t_2^2d_k^2}{\eta_e^2-\eta_o^2}\cos{k_2}d^2k,
    \\
    & \tilde{g}_{2ai}=-\frac{1}{4\pi^2}\int \frac{\eta_ed_k+2m^2d_k}{\eta_e^2-\eta_o^2}\sin{k_2}d^2k,
    \\
    & g_{31}=-\frac{1}{4\pi^2}\int
    \text{Re}[h_{k}]
    \frac{(1-\cos(k_1+k_2))\eta_e}{\eta_e^2-\eta_o^2}d^2k,
\end{align}
where $\eta_e$, $\eta_o$ are given by
\begin{align}
    &\eta_e=-m^2-t_2^2d_k^2- |h_k|^2, 
    &\eta_o=2 m t_2 d_k     
\end{align}
Note that since ${g}_{31}<0$, $0<\theta_0<\pi$. Therefore, $\sin({\theta_0}/{2})>0$ and $\cos({\theta_0}/{2})>0$.

Eq.~\ref{eq_hc_haldane} defines the two critical points of zero energy crossings. 
Each crossing involves two bound states coming from the $\alpha=+$ and $\alpha=-$ sectors. The states are computed as the eigenvectors of Eq.~\ref{eq_tmatrix_haldane}, giving
\begin{align}
    \psi_{\alpha}^{ V_R}({\vec r},S)=f_{\beta\alpha}^Se^{i\beta \ell\phi_d}
    \label{eq_haldane_psi_d}
\end{align}
with $f^A_{\beta\alpha}=\cos\left(\frac{\theta_0}{2}+\frac{\alpha\beta-1}{4}\pi\right)$, $f^B_{\beta\alpha}=\sin\left(\frac{\theta_0}{2}-\frac{\alpha\beta-1}{4}\pi\right)$, and  $\ell=(3-\alpha\beta)/2$. 
(The inverted relation is $\alpha_\ell=\beta(3-2\ell)$.) 
Here $\phi_d$ is the polar angle of each of the six sites $\vec{r}$ of the defect hexagon $\hexagon'$ (with its center as the origin). Since $f^S_{\beta\alpha}>0$, the phase of the wavefunctions show angular momenta $\beta \ell$. 
Note that for the special particle-hole and sublattice symmetric limit of $m=0$, we have $f^A_{\beta\alpha}=f^B_{\beta\alpha}=\frac{1}{\sqrt{2}}$ and these states become  the eigenstates of $C_6$, as expected from the $C_6$ rotation symmetry of $H_0+V_R$. For $m\neq0$, the $C_6$ symmtery is broken by the sublattice imbalance, but still  the wavefunction shows the phase winding $\beta \ell$, along with a sublattice modulation of its magnitude. 

Since $G_0$ obeys the threefold rotation symmetries of the model, the extended part of the wavefunction has the same $C_3$ phase winding as $\psi_{\alpha}^{ V_R}$.
We now argue that it can be written as Eq.~\ref{eq_dbar_mzero} with the required sublattice properties of $g^S_{\beta\alpha}(\rho)$. 
For the sublattice symmetric limit of $m=0$, $g^S_{\beta\alpha}=g_{\beta\alpha}$ becomes sublattice independent implementing $C_6$ symmetry. Now consider the generic $m$ case.
On the defect sites Eq.~\ref{eq_haldane_psi_d} together with $f^S_{\beta\alpha}>0$ show that although the amplitudes differ between sublattices, the phase remains the same.
To see that this property is preserved by the extended part of the state, 
we compute the expectation value of the Haldane model mass term  in the  wavefunction of Eq.~\ref{eq_dbar_mzero}: $\langle m\sigma^z\rangle=m\sum_{r}\left(|g^A_{\beta\alpha}\left(\rho\right)|^2-|g^B_{\beta\alpha}\left(\rho\right)|^2\right)$. No relative phase  appears; the mass term does not favor phase differences. The remaining Haldane model terms, which have $C_6$ symmetry, produce the $C_6$ phase winding as for $m=0$.
Therefore adding the mass term should only change the  amplitudes of $g_{\beta\alpha}^S(\rho)$ between sublattices, 
 resulting in Eq.~\ref{eq_dbar_mzero}.

\subsubsection{Effective low-energy model.}
Computing $t_{R R'}^{\ell \ell'}$ is relatively simple for the QWZ model. 
Using its $G_0$ given above, we have $s^{11}_{RR'}=s^{22}_{RR'}=g_{11}(R-R')$
which at large distances becomes $\rightarrow g_0(|R-R'|/\xi)$ with $\xi$ the correlation length of $H_0$ and $g_0(x)=\sqrt{\frac{\pi}{2x}}e^{-x}$. The off diagonal element has a similar expression in terms of $g_{12}$ and they all become equal at long distance, with $s^{11}_{RR'}=s^{12}_{RR'}=g_0(|R-R'|/\xi)$.

 For the Haldane model the computation involves a sum over the sites in the two defects. 
Generally we note that  with generic sublattice imbalance mass term $m$, the amplitudes of these functions $s$ become $\ell$ orbital dependent, however the  sign and phase remains independent of the orbitals.

 To do the computation we note that using $\alpha_\ell=\beta(3-2\ell)$ we can express
$V_R|\psi^{V_R}_{\ell}\rangle=\sqrt{3}h\alpha_\ell|\psi^{V_R}_{\ell}\rangle$. 
 Then the issue is the sum over sites, which can be done in a multipole expansion. The zeroth order (monopole) term describes a defect with vanishing size, which pulls out $G_0$, leaving a sum over the exponential phases which identically vanishes. The first order (dipole) term gives the leading contribution. With this term 
 $t^{ll'}_{RR'}$ becomes
\begin{align}
    \alpha_\ell\alpha_{\ell'}(-1)^{\ell'}3h^2 f_{\beta\alpha_\ell}^s ~
    f_{\beta\alpha_{\ell'}}^{s'} ~ G_0(\vec{\tilde{r}}-\vec{ \tilde{r}}';s,s') ~ e^{i\beta (\ell'-\ell)\theta_{RR'}}
    \label{eq_hopping_hexa}
\end{align}
giving the effective Hamiltonian of Eq.~\ref{eq_eff_hopping}.
See  SI for details and for the next order correction.

\subsubsection{Disorder averaged TAI transitions.}

Including impurities the full disorder averaged Green's function is
\begin{align}
    \langle G(\omega)\rangle&=\left\langle\frac{1}{\omega-H}\right\rangle=G_{\rm clean}(\omega)+G_{\rm clean}(\omega)\langle\Sigma(\omega)\rangle \langle G(\omega)\rangle,
\end{align}
where $G_{\rm clean}$ is the Green's function without impurities, $
    G_{\rm clean}(\omega)=1/(\omega-H_0)$ (while $G_0=G_{\rm clean}(0)$)
and  $\langle\Sigma\rangle$ is the disorder averaged self-energy,
\begin{align}
    \langle\Sigma(\omega)\rangle=\langle V\rangle+\langle VG_{\rm clean}(\omega)V\rangle_c+\cdots
\end{align}
with $V=\sum_R V_R$ and $\langle\cdots\rangle_c$ denotes the connected correlations. 
This disorder averaged self energy renormalizes the bulk Hamiltonian to $H_0+\langle\Sigma\rangle$.  

In our models, the disorder forms a Bernoulli distribution where probability of a site (or second neighbor bond) to have defect is $p=n_d$ ($p=2n_d$) for QWZ (Haldane) model. (Haldane model has two $t_2$ bonds per site.)

For the QWZ model we have, to order $h^2$,
\begin{align}
 \langle V\rangle&=-\mathcal{N} n_d h\sigma^z,\\
     \langle V G_{\rm clean}  V\rangle_c
     &=
    \sum_{RR'}\langle V_R^2\rangle_c\delta_{RR'}G_{\rm clean}^{RR}
    \\
    &=\mathcal{N}\langle V_R^2\rangle_c \int \frac{d^2k}{(2\pi)^2}G_{\rm clean}(\omega=0,k)
    \\
    &=\mathcal{N}n_d(1-n_d)h^2g_{\rm onsite}\sigma^z, 
\end{align}
where $\mathcal{N}$ is the total number of sites and $\delta_{RR'}$  appears because the disorder distribution is uncorrelated. The self-energy renormalizes $u\rightarrow u+\mathcal{N}^{-1}\langle\Sigma\rangle$.

For the Haldane model, we similarly have 
\begin{align}
    \langle V\rangle&=\sum_{\langle\langle ij\rangle\rangle}(2n_d)(-ih)c_i^\dagger c_j+\text{H.c.}. %
    \\
    \langle V G_{\rm clean}  V\rangle_c&=\mathcal{N}\sum_{R}\langle V_R^2 \rangle_c G_{\rm clean}\\
    &=\mathcal{N}\sum_{r_1r_2r_3r_4}\langle V_R^{r_1r_2} V_R^{r_3r_4}\rangle _cG^{r_2r_3}_{\rm clean}\\
    &=\mathcal{N}n_d(1{-}n_d)\sum_{r_1r_2r_3r_4} t_R^{r_1r_2} t_R^{r_3r_4} G^{r_2r_3}_{\rm clean}
    \label{eq_second order}
\end{align}
Here  $r_i$ denote sites in a hexagon defect, and $t_{R}^{r_ir_j}$ is the hopping amplitude between the two sites. The expression \ref{eq_second order} renormalizes both the hopping parameters and onsite mass $m$. 

We proceed with computing this $h^2$ term. 
To compute the renormalization of the second neighbor hopping, we need to look at the $\langle VG_{\rm clean}V\rangle_{r_1r_4} $ element where $r_1$ and $r_4$ are two endpoints of a second neighbor bond, say two $A$ points of the defect hexagon at $R$. This forces $r_2$ and $r_3$ to be on the other $A$ sublattice point on the hexagon. So $G_{\rm clean}^{r_2r_3}$ is the onsite Green's function, which is  real.  Since $t_{R}^{r_ir_j}$ is purely imaginary, this matrix element is real and generates purely real second neighbor hopping $-n_d(1-n_d)g^{S}_{\rm onsite}h^2$ (where $g^{S}_{\rm onsite}$ is the onsite Green's function for $S$ sublattice) without modifying $t_2$.
Similarly $m$ renormalizes to
$
    m\rightarrow m+2n_d(1-n_d)h^2\left(g_{\rm onsite }^{A}-g_{\rm onsite }^{B}+{\rm Re}\left(g_{\rm 2nd}^{A}-g_{\rm 2nd }^{B}\right)\right),
$
where $g^{S}_{\rm 2nd}$ is the second neighbor $G_{\rm clean}$ for sublattice $S$.

\subsubsection{Numerical computations.} 

In the computations for Fig. \ref{fig:bott_ipr} we used a torus with 1024 sites and 31 defects distributed randomly with no two defects touching.
The  Bott index   is computed by~\cite{loring_disordered_2011}:
\begin{align}
    \mathcal{B}_0= {\rm Re}\left(\frac{1}{2\pi i} {\rm tr}\left(\log\left(V_{x_1}  V_{x_2} V_{x_1}^\dagger  V_{x_2}^\dagger\right)\right)\right),
    \label{eq_bott}
\end{align}
\begin{align}
    V_{x_j}=\mathcal{P}_0  \exp\left( \frac{2\pi i}{L_{j}} x_j\right)\mathcal{P}_0+\mathcal{Q}_0.
    \label{eq_V-bott}
\end{align}
Here $\mathcal{P}_0$ is the projector onto the occupied states and  $\mathcal{Q}_0=1-\mathcal{P}_0$ is the projector onto unoccupied states.
To investigate the energy scales involved in producing the Bott index, we also compute the effects of modifying the occupancy of states in the projector $\mathcal{P}_0$,
using a  Fermi-Dirac distribution at a given temperature.

\bibliography{references}

\end{document}